\documentclass[11pt]{iopart}
\usepackage{acronym}
\usepackage{amssymb}
\usepackage{xcolor}
\usepackage{graphicx}
\usepackage{url}
\usepackage{ulem}[normal]

\newcommand{\newtxt}[1]{{#1}}

\newcommand{\eqref}[1]{Eq.~(\ref{#1})}

\begin{document}

\title{Do black holes remember what they are made of?}

\author{Harshraj Bandyopadhyay$^1$, David Radice$^{1,2,3}$\footnote{Alfred P.~Sloan Fellow}, Aviral Prakash$^{1,2}$, Arnab Dhani$^{4}$, 
Domenico Logoteta$^{5,6}$, Albino Perego$^{7,8}$, and Rahul Kashyap$^{1,2}$}

\address{${}^1$ Department of Physics, The Pennsylvania State University, University Park, PA 16802}
\address{${}^2$ Institute for Gravitation \& the Cosmos, The Pennsylvania State University, University Park, PA 16802}
\address{${}^3$ Department of Astronomy \& Astrophysics, The Pennsylvania State University,University Park, PA 16802}
\address{${}^4$ Max Planck Institute for Gravitational Physics (Albert Einstein Institute), Am Mühlenberg 1, Potsdam 14476, Germany}
\address{${}^5$ Dipartimento di Fisica, Universit\`{a} di Pisa, Largo B.  Pontecorvo, 3 I-56127 Pisa, Italy}
\address{${}^6$ INFN, Sezione di Pisa, Largo B. Pontecorvo, 3 I-56127 Pisa, Italy}
\address{${}^7$ Dipartimento di Fisica, Università di Trento, Via Sommarive 14, 38123 Trento, Italy}
\address{${}^8$ INFN-TIFPA, Trento Institute for Fundamental Physics and Applications, Via Sommarive 14, I-38123 Trento, Italy}

\ead{dur566@psu.edu}

\begin{abstract}
We study the ringdown signal of black holes formed in prompt-collapse binary neutron star mergers. We analyze data from \newtxt{$47$} numerical relativity simulations. We show that the $(\ell=2,m=2)$ and $(\ell=2,m=1)$ multipoles of the gravitational wave signal are well fitted by decaying damped exponentials, as predicted by black-hole perturbation theory. We show that the ratio of the amplitude in the two modes depends on the progenitor binary mass ratio $q$ and reduced tidal parameter $\tilde\Lambda$. Unfortunately, the numerical uncertainty in our data is too large to fully quantify this dependency. If confirmed, these results will enable novel tests of general relativity in the presence of matter with next-generation gravitational-wave observatories.
\end{abstract}

\section{Introduction}
The \ac{GW} signal from binary \ac{NS} mergers spans a wide frequency range \cite{Radice:2020ddv}. Binary \ac{NS} systems can be detected by ground-based observatories, such as LIGO \cite{LIGOScientific:2014pky}, Virgo \cite{VIRGO:2014yos}, and KAGRA \cite{KAGRA:2020tym}, when their orbital period is less than ${\sim}0.1$~seconds, corresponding to \ac{GW} frequencies of ${\sim}20\ {\rm Hz}$. As the stars inspiral the orbital frequency increases. Tidal effects start to become important at frequencies of a few hundreds Hz \cite{Flanagan:2007ix, Hinderer:2009ca, Damour:2009wj, Damour:2012yf}, enabling \ac{GW} observation to constrain the structure of \acp{NS} and their \ac{EOS} \cite{LIGOScientific:2017fdd, De:2018uhw, LIGOScientific:2018hze, LIGOScientific:2018cki}. The stars come into contact at a frequency between 1 and 2 kHz \cite{Damour:2009wj}. If the binary does not result in the immediate formation of a \ac{BH} \cite{Shibata:2005ss, Hotokezaka:2011dh, Bauswein:2013jpa, Zappa:2017xba, Agathos:2019sah, Koppel:2019pys, Bauswein:2020aag, Bauswein:2020xlt, Perego:2021mkd, Kashyap:2021wzs, Kolsch:2021lub, Cokluk:2023xio}, a \ac{RMNS} is at least temporarily formed \cite{Baumgarte:1999cq, Rosswog:2001fh, Shibata:2006nm, Baiotti:2008ra, Sekiguchi:2011zd, Hotokezaka:2013iia,  Palenzuela:2015dqa, Kiuchi:2017zzg, Radice:2017zta, Kiuchi:2022nin, Radice:2018xqa, Palenzuela:2021gdo, Palenzuela:2022kqk, Radice:2023zlw}. Such remnants are efficient emitters of \acp{GW} in the frequency range of $1{-}6$~kHz \cite{Shibata:2005xz, Hotokezaka:2011dh, Bauswein:2013jpa, Takami:2014zpa, Bernuzzi:2015rla, Radice:2016rys, Most:2018eaw, Most:2019onn, Bauswein:2018bma, Prakash:2021wpz, Raithel:2021hye, Fields:2023bhs, Prakash:2023afe}. If \ac{BH} formation occurs, the remnant \ac{BH} rings down as it relaxes to a Kerr \ac{BH} producing \acp{GW} at even higher frequencies of $6{-}8$~kHz \cite{Baiotti:2005vi, Zhang:2020qlh, Dhani:2023ijt}.

The ringdown signal from newly formed \acp{BH} is expected to be described by \ac{BH} perturbation theory \cite{Berti:2009kk}. Accordingly, the signal can be described as a superposition of discrete \acp{QNM}, which decay exponentially in time, and a tail signal that decays as a power law. The predictions of \ac{BH} perturbation theory have been experimentally confirmed in the case of the binary \ac{BH} mergers GW150914 \cite{LIGOScientific:2016lio, Isi:2019aib} and GW190521 \cite{Capano:2021etf, LIGOScientific:2020tif, Siegel:2023lxl}\newtxt{, as well as for 20 other binary \ac{BH} events \cite{LIGOScientific:2020tif, LIGOScientific:2021sio}}. \newtxt{While there is consensus on the fact that the ringdown signal for all these events is consistent with the predictions of general relativity, there is a controversy with the interpretation of possible higher-order modes in these signals} \cite{Carullo:2019flw, Cotesta:2022pci, Finch:2022ynt, Baibhav:2023clw, Zhu:2023mzv, Isi:2023nif}. According to perturbation theory, the oscillation frequencies and damping times of the \acp{QNM} depend only on the \ac{BH} mass and spin. However, the ``initial'' amplitudes of the different modes depends on how the \ac{BH} is formed. In the case of binary \ac{BH} mergers, several works pointed out that the amplitudes encode information of the binary prior to merger \cite{Kamaretsos:2012bs, Damour:2014yha, Healy:2014eua, London:2014cma, London:2018gaq, Berti:2018vdi, Baibhav:2018rfk, Cotesta:2018fcv, Lim:2019xrb, Borhanian:2019kxt, Hughes:2019zmt, Forteza:2022tgq, Zhu:2023fnf}. This can enable model-independent tests of general relativity by checking for consistence between inspiral and post-merger signals.

In a recent work, Dhani and collaborators \cite{Dhani:2023ijt} showed that \acp{QNM} from binary \ac{NS} mergers producing \acp{BH} could also be detected by next generation \ac{GW} experiments, such as the Einstein Telescope \cite{Punturo:2010zz} and Cosmic Explorer \cite{Reitze:2019iox}. This motivates us to check whether or not the ringdown signal of \acp{BH} formed by such mergers also encodes information about the binary prior to merger\footnote{\newtxt{This information might also be encoded in the ``pre-Hawking radiation'' emitted during the collapse \cite{Dai:2016sls}.}}. This is particularly important because there are alternative theories of gravity, such as certain classes of scalar-tensor theories, that differ from general relativity only if matter is present, e.g.,~\cite{Barausse:2012da, Sagunski:2017nzb, East:2022rqi, Bezares:2021dma}. As such, \ac{NS} mergers provide the only way to constrain such theories. In this work, we analyze results from \newtxt{47} numerical relativity simulations. The data suggests that the $(\ell=2,m=1)$ and $(\ell=2,m=2)$ modes might depend on the mass ratio and tidal parameters of the binary. However, the quality of the current numerical relativity data is insufficient to arrive to firm conclusions. If these results could be confirmed with more accurate simulations, then this would open a new avenue for testing general relativity with next-generation ground-based experiments. The rest of this manuscript is organized as follows. In Section \ref{sec:methods} we describe our simulations and the approach used to extract the \ac{QNM} amplitudes from the data. We discuss our main results in Section \ref{sec:results}. Finally, Section \ref{sec:conclusions} is dedicated to discussion and conclusions.

\section{Methods}
\label{sec:methods}

\subsection{Numerical Relativity Simulations}

\begin{figure}
    \centering
    \includegraphics{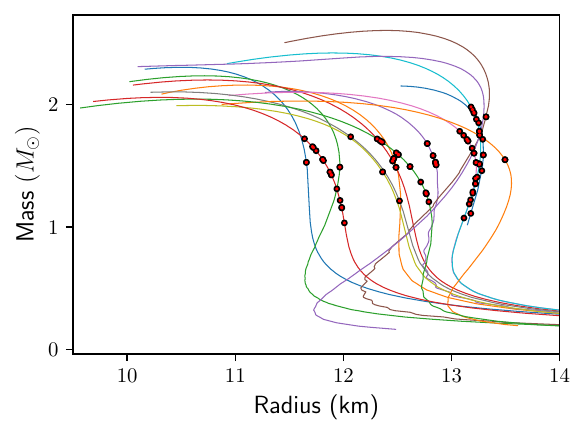}
    \caption{Mass radius curves for the EOSs employed for this work. The red dots mark the specific configurations considered.}
    \label{fig:mass.radius}
\end{figure}

We consider a set of \newtxt{47} numerical relativity simulations spanning 15 \acp{EOS}. All simulations have been performed with the \texttt{WhiskyTHC} code \cite{Radice:2012cu, Radice:2013hxh, Radice:2013xpa} starting from quasi-circular configurations created with \texttt{Lorene} \cite{Gourgoulhon:2000nn}. Our simulations have a spatial resolution of $0.125\, G M_\odot / c^2 \simeq 185\, {\rm m}$ which we refer to as the standard resolution. Additional 18 simulations have been performed at a lower resolution of $0.167\,  G M_\odot / c^2 \simeq 246\, {\rm m}$, to estimate the numerical errors. These simulations have been presented in Refs.~\cite{Kashyap:2021wzs, Perego:2021mkd} to which we refer for additional details. To have a homogeneous dataset, we only consider systems that underwent prompt \ac{BH} formation. For the purpose of this study, we define a prompt \ac{BH} formation to take place when the lapse function ($\alpha$) after merger decreases monotonically with time until \ac{BH} formation (no bounce). Our dataset comprises total binary masses $M \in [2.75,3.8]\, M_\odot$ and mass ratios $q \in [0.57, 1]$.

Figure~\ref{fig:mass.radius} shows the span of the considered \acp{EOS}. \newtxt{They cover the range of radii and masses compatible with astrophysical constraints, e.g.,~\cite{Breschi:2024qlc}. The \acp{EOS} are diverse not only in the macroscopic properties of \acp{NS} they predict, but also in the nuclear theory techniques and microphysics they model. We consider six relativistic mean-field theory models: Big-Apple \cite{Fattoyev:2020cws}, BHB$\Lambda\phi$ \cite{Banik:2014qja}, HS(DD2) \cite{Typel:2009sy, Hempel:2009mc}, DD2qG (Logoteta et al., in prep), H4 \cite{Glendenning:1991es, Lackey:2005tk, Read:2008iy}, and SFHo \cite{Steiner:2012rk}; six models employing Skyrme-type interactions: LS220 \cite{Lattimer:1991nc} and five variants of the SRO \ac{EOS}  \cite{Schneider:2017tfi, Schneider:2019shi}; two microphysical Hartree-Fock models: BLh \cite{Bombaci:2018ksa, Logoteta:2020yxf} and BLQ \cite{Prakash:2021wpz}; and one purely phenomenological piecewise polytropic model from Refs.~\cite{Godzieba:2020tjn, Kashyap:2021wzs}. Of these \acp{EOS}, two include deconfined quarks, BLQ and DD2qG, two include hyperons, BHB$\Lambda\phi$ and H4, while the others are purely nucleonic.}

\subsection{Data Preparation}
\ac{GW} data is extracted from the simulations using the Newman-Penrose formalism \cite{Bishop:2016lgv}. Accordingly, we record the complex scalar $\Psi_4$ extracted on a coordinate sphere with radius $R = 400\, G M_\odot / c^2 \simeq 591\, {\rm km}$. We then recover the complex strain $h = h_+ - i h_\times$ from $\Psi_4$ using the fixed-frequency integration method \cite{Reisswig:2010di}.

We consider two of the spherical-harmonic modes of the ringdown signal: $(\ell=2, m=2)$ and $(\ell=2, m=1)$. Each mode is modeled as a damped sinusoid with the ansatz,
\begin{equation}\label{eq:ansatz}
    h_{\ell m}^{\rm fit}(t) = A_{\ell m} e^{-i\, \omega_{\ell m}\, (t - t_0)}, 
\end{equation}
where $A_{\ell m}$ and $\omega_{\ell m}$ are the complex amplitude and frequency for each mode and $t_0$ is a reference time, specified below. The fit is performed using the data in the time window $[t_i, t_e]$, where $t_i$ is varied over $[0,496] \, G M_\odot/ c^3 \simeq [0, 2.44]\, {\rm ms}$ and $t_e$ is fixed to $500\, G M_\odot/ c^3 \simeq 2.46\, {\rm ms}$. $t=0$ is taken to be the time of merger, conventionally defined as the time at which the ($l=2$, $m=2$) mode amplitude peaks. For each fixed value of $t_i$, and separately for $(\ell=2, m=2)$ and $(\ell=2, m=1)$, we obtain $A_{\ell m}$, $\omega_{\ell m}$ by minimizing the mismatch
\begin{equation}
    \mathcal{M} = 1 - \frac{\langle h, h^{\rm fit} \rangle}{\sqrt{\langle h^{\rm fit}, h^{\rm fit} \rangle \langle h, h \rangle}}\,,
\end{equation}
where
\begin{equation}
    \langle u, v \rangle = \int_{t_i}^{t_e} u(t)\, v^\ast(t)\, {\rm d}t\,.
\end{equation}
To speedup the calculation, we use the best fitting values from the previous $t_i$ as initial guesses for each nonlinear least square fit after the first. This procedure yields a range of values $A_{\ell,m}(t_i)$, $\omega_{\ell,m}(t_i)$, and fits for each time.

\begin{figure}
    \centering
    \includegraphics{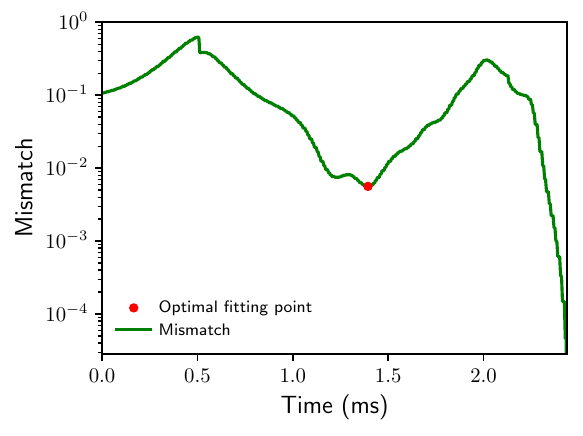}
    \caption{Mismatch as a function of the fitting starting time (green curve) and optimal fitting point (red point). The results are for a binary with $M_1 = 1.929\, M_\odot$, $M_2 = 1.351\, M_\odot$ simulated with the HS(DD2) \ac{EOS}.}
    \label{fig:mismatch}
\end{figure}

To obtain the optimal values for $A_{\ell m}$ and $\omega_{\ell m}$ for each simulation, we follow a procedure adopted in many other works, e.g.,~Refs.~\cite{Mourier:2020mwa, Forteza:2021wfq, Dhani:2021vac, MaganaZertuche:2021syq}. Namely, we consider the set of minimal mismatches $\left\{ \mathcal{M}(t_i) \right\}$ obtained when fitting over the windows $[t_i, t_e]$ and select the $t_i$ for which this quantity attains its first minimum. For example, $\{ \mathcal{M}(t_i) \}$ for the $(\ell = 2, m = 2)$ mode is shown in Fig.~\ref{fig:mismatch} for a binary with $M_1 = 1.929\, M_\odot$, $M_2 = 1.351\, M_\odot$ simulated with the HS(DD2) \ac{EOS} \cite{Typel:2009sy, Hempel:2009mc} at standard resolution, which we take as fiducial. The results for other binaries are qualitatively similar. We find that the mismatch has a clear minimum ${\sim}1.5$~ms after merger. The mismatch is also small for much larger $t_i$'s, but this is because the model and the strain have become roughly constant. For this reason, we take the minimum marked with a red dot as ``optimal'' and record for each binary the fitted (complex) amplitudes $A_{21}$ and $A_{22}$, and frequencies, $\omega_{21}$ and $\omega_{22}$, corresponding to this starting time. Note that the optimal starting times for the $(\ell=2,m=2)$ and $(\ell=2,m=1)$ modes typically differ by ${\sim}14\, G M_\odot/c^3 \simeq 70 \ {\rm \mu s}$. In order to meaningfully compare the amplitude in the two modes we are considering, we fix $t_0$ to be the optimal fitting time for the $(\ell=2,m=2)$ mode. In other words, $A_{21}/A_{22}$ is equal to $|h_{21}|/|h_{22}|$ at time $t_0$. Finally, we remark that we do not impose any relation between the complex frequencies. This procedure is repeated for all binaries. In each case, the correct minimum is first identified by visual inspection and then located by sampling $\mathcal{M}$ with $\Delta t_i$ of $1 \, G M_\odot/ c^3$. We obtain minimum mismatches $\mathcal{M}$ in the range $1.4 \times 10^{-3}$ to $0.48$ with average $1.7\times 10^{-2}$ for the $(\ell=2,m=2)$ mode, and in the range $9.5 \times 10^{-3}$ to $0.12$ with average $3.0\times 10^{-2}$ for the $(\ell=2,m=1)$ mode, for the binaries included in this study.

\begin{figure}
    \centering
    \includegraphics{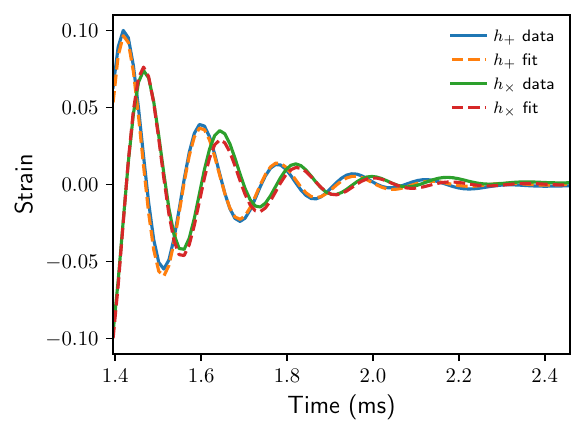}
    \caption{Real and imaginary part of the $(\ell=2,m=2)$ strain and optimal fit for a binary with $M_1 = 1.929\, M_\odot$, $M_2 = 1.351\, M_\odot$ simulated with the HS(DD2) \ac{EOS}. The starting point of the fit is shown in Fig.~\ref{fig:mismatch}.}
    \label{fig:strain}
\end{figure}

The $(\ell=2,m=2)$ fit of the strain resulting from the fitting procedure for our fiducial binary are shown in Fig.~\ref{fig:strain}. Overall, we find that our ansatz fits the numerical relativity data well, suggesting that the late-time \ac{GW} signal is indeed a \ac{BH} \ac{QNM}. However, at late times the numerical relativity signal does not decay to zero, but typically plateaus and exhibits further oscillations. This is better seen in log-scale, as shown in Fig.~\ref{fig:amplitude}, again for our fiducial binary. We interpret this as numerical noise. The data for the $(\ell=2,m=1)$ mode is qualitatively similar for binaries with $q \neq 1$. In the $q=1$ case, we do not expect the $(\ell=2,m=1)$ mode to be excited in the $q=1$ case. In practice, this mode can be excited due to symmetry breaking during merger \cite{Paschalidis:2015mla, Radice:2016gym, Lehner:2016wjg, Espino:2023llj}. However, the resulting amplitudes are small and are not well fitted with the QNM ansatz of~\eqref{eq:ansatz}. As such, we do not compute $A_{21}$ for the equal mass binaries. As discussed below, all data considered in this study shows good agreement with the predictions from \ac{BH} perturbation theory over at least two full \ac{GW} cycles.

\subsection{Data Quality}\label{sec:data.quality}

\begin{figure}
    \centering
    \includegraphics{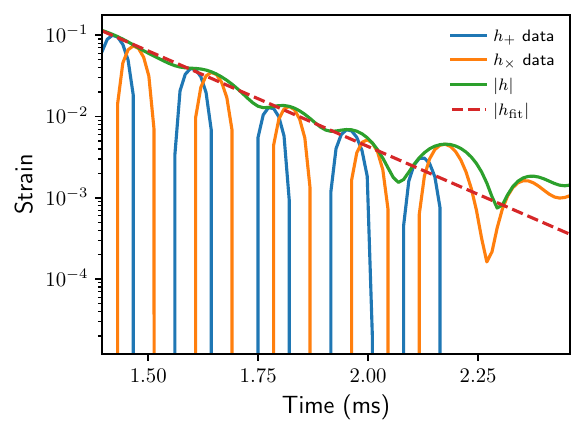}
    \caption{Same as Fig.~\ref{fig:strain}, but in log scale. Here, we also show the overall amplitude of the signal $|h|=\sqrt{h_+^2 + h_\times^2}$ and of its fit $|h_{\rm fit}|$.}
    \label{fig:amplitude}
\end{figure}

After having constructed the optimal fit for all simulations, we manually inspect all data using diagnostic plots similar to that in Fig.~\ref{fig:amplitude}. We remove all models for which the numerical relativity data shows less than two full \ac{GW} cycles of close agreement with our damped sinusoidal fit. This selection procedure is also performed for the $(\ell=2,m=1)$ mode. Out of an initial set of 58 binaries at standard resolution, we remove \newtxt{11} cases ending with the \newtxt{47} binaries showed in Fig.~\ref{fig:mass.radius}.

We estimate finite-resolution uncertainties from 14 unequal-mass and 4 equal-mass binaries that have been simulated at two resolutions: standard and low resolution. The relative error in $A_{21}/A_{22}$ ($A_{22}$) range from 1.7\% (0.2\%) to 104\% (66\%). We take the mean value of the error as typical relative uncertainty: 37\% and 33\% for $A_{21}/A_{22}$ and $A_{22}$, respectively. We believe our error estimate to be very conservative, because we have only simulated at two resolutions the binaries close to the threshold for prompt collapse. We expect the errors to be largest for these binaries.

\section{Results}
\label{sec:results}

\begin{figure}
    \centering
    \includegraphics{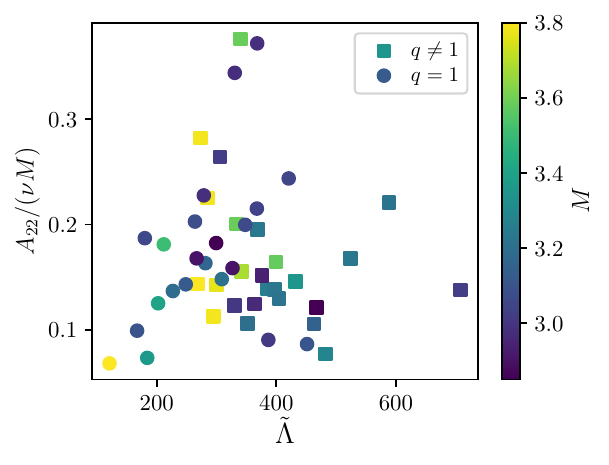}
    \caption{Normalized amplitude $A_{22}/(\nu M)$, $\nu$ being the symmetric mass ratio $M_A M_B / M^2$ and $M = M_A + M_B$ being the reduced mass, as a function of the reduced tidal deformability $\tilde\Lambda$. The color encodes the total mass $M = M_A + M_B$. Squares (circles) denote equal (unequal) mass binaries.}
    \label{fig:A22}
\end{figure}

Figure~\ref{fig:A22} shows the fitted amplitude of the $(\ell=2,m=2)$ mode in the ringdown signal normalized by the total mass $M = M_A + M_B$ and symmetric mass ratio $\nu = M_A M_B / M$, $A_{22}/(M \nu)$, as a function of the reduced tidal parameter
\begin{equation}
    \tilde{\Lambda} = \frac{16}{13} \left[\frac{(M_A + 12 M_B) M_A^4  \tilde\Lambda_A }{(M_A + M_B)^5} + (A \leftrightarrow B)\right]\,,
\end{equation}
where $\Lambda_{A,B}$ are the dimensionless quadrupolar tidal parameters (or tidal polarizability coefficients) of each of the stars in the binary. The figure also encodes the total mass of the binary with color. Overall, we do not find evidence for a correlation between $A_{22}/(\nu M)$, $\tilde\Lambda$, $M$, or $q$. For reference, we estimated $A_{22}/(\nu M)$ for an equal-mass, nonspinning binary \ac{BH} value merger to be $0.496$ using the publicly available data from the SXS catalog \texttt{SXS:BBH:0066} \cite{Boyle:2019kee}. Note that the value we are quoting is the amplitude of the signal at the best fitting time, following the convention we are using in this work, and not at merger, as often done in the binary \ac{BH} literature.

\begin{figure}
    \centering
    \includegraphics{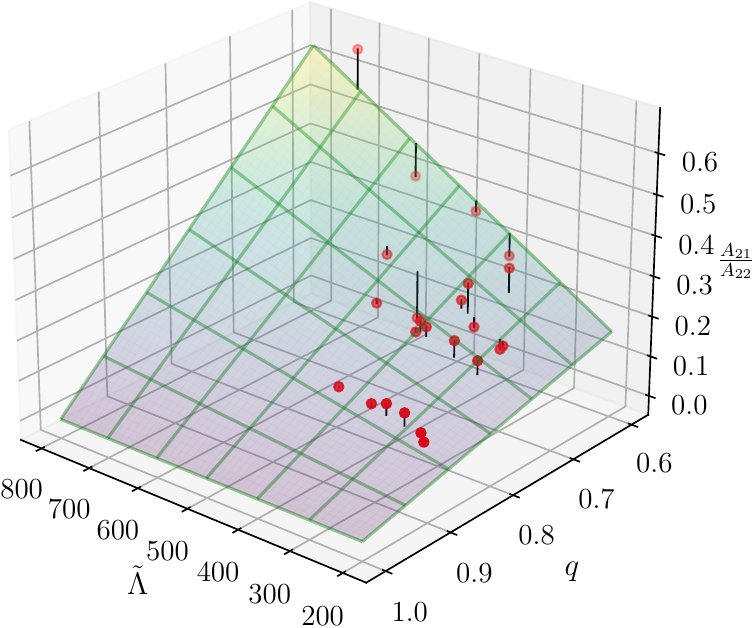}
    \caption{Ratio of the amplitude of the $(\ell=2,m=1)$ and $(\ell=2,m=2)$ QNM modes, $A_{21}/A_{22}$, as a function of binary mass ratio $q$ and tidal parameter $\tilde{\Lambda}$. The surface shows the tentative fit of \eqref{eq:fit}. The vertical lines show the distance between the data and the fit. Residuals are also shown in Fig.~\ref{fig:residuals}.}
    \label{fig:A21_o_A22}
\end{figure}

Figure~\ref{fig:A21_o_A22} shows the ratio of the amplitudes of the $(\ell=2,m=1)$ and $(\ell=2,m=2)$ modes, $A_{21}/A_{22}$, as a function of $q$ and $\tilde\Lambda$. We find that this ratio depends on both $q$ and $\tilde\Lambda$. First, for $q=1$ the amplitude of the $(\ell=2,m=1)$ mode is close to zero and, as previously discussed we cannot reliably measure $A_{21}$. This is because the binary is symmetric with respect to rotations of $180^\circ$ about the $z$-axis (normal to the orbital plane). Note that such symmetry is known to be broken in the postmerger, as the result of the one-armed spiral instability \cite{Paschalidis:2015mla, Radice:2016gym, Lehner:2016wjg, Espino:2023llj}. However, this instability does not have a chance to fully develop for the prompt \ac{BH} formation binaries considered in this study. Moreover, we find that in general $A_{21}/A_{22}$ grows with mass asymmetry and tidal deformability. We quantify this by fitting the data with the ansatz
\begin{equation}\label{eq:fit}
    (1 - q) \left(\frac{a}{1+q} + b\, \tilde{\Lambda}\right)\,,
\end{equation}
where $a$ and $b$ are determined with an ordinary least square fit to be:
\begin{equation}
    a = 3.956 \times 10^{-2}, \qquad
    b = 1.952 \times 10^{-3}\,.
\end{equation}
Our ansatz \eqref{eq:fit} is a generalization the fit originally proposed in Ref.~\cite{Kamaretsos:2012bs} for binary \ac{BH} merger data. The fit $R^2$ is $0.86$. The fit is also shown as the surface in Fig.~\ref{fig:A21_o_A22}. \newtxt{It is interesting to compare our results to the case of binary \acp{BH}, which would formally correspond to the limit $\tilde\Lambda \to 0$.} For comparison, we refit this expression to nonspinning binary \ac{BH} data from the SXS catalog \cite{Boyle:2019kee} with $q$ between $1$ and $10$  (and $\tilde\Lambda = 0$) to obtain $a' = 0.216$. Note that this coefficient is different from that reported in  Ref.~\cite{Kamaretsos:2012bs}, because we define the ratio $A_{21}/A_{22}$ at $t_0$ instead of at the merger time. That said, because $\tilde\Lambda$ and $q$ are strongly correlated, $a$ and $a'$ cannot be directly compared. If we fit the binary \ac{NS} data with the same ansatz as the binary \ac{BH} data, that is if we neglect $\tilde\Lambda$, we obtain $a''= 1.454$ with a lower $R^2 = 0.61$. Overall, our results suggest that matter effects strongly amplify $A_{21}$ compared to the expectation for binary \acp{BH}. \newtxt{In other words, the limit $\tilde\Lambda \to 0$ appears to be discontinuous.} We remark that there is some scatter in the data and the numerical uncertainties in $A_{21}/A_{22}$ are large. As such, while we believe that the correlations we are measuring are robust, our data might not yet be sufficiently accurate do determine its exact dependency.

\begin{figure}
    \centering
    \includegraphics{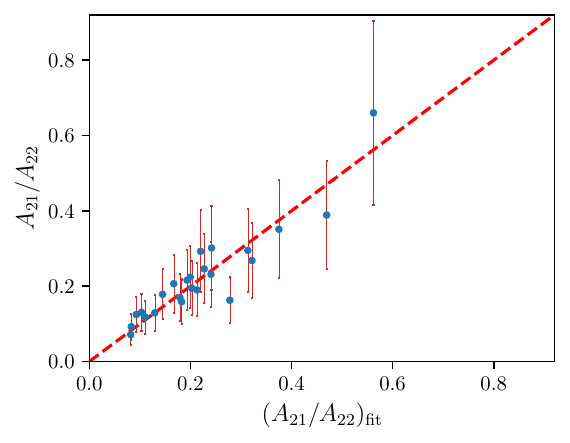}
    \caption{Ratio of the amplitudes of the $(\ell=2,m=1)$ and $(\ell=2,m=2)$ modes, $A_{21}/A_{22}$, vs the fit $(A_{21}/A_{22})_{\rm fit}$ from \eqref{eq:fit}.}
    \label{fig:residuals}
\end{figure}

Figure~\ref{fig:residuals} shows the ratio of the amplitudes of the $(\ell=2,m=1)$ and $(\ell=2,m=2)$ modes as a function of  the fit $(A_{21}/A_{22})_{\rm fit} = (1 - q) \big[a/(1+q) + b\, \tilde{\Lambda}\big]$. The error bars show the 37\% variation on the data (see Sec.~\ref{sec:data.quality}). We find that the data is broadly consistent with the fit, once error bars are taken into account. There is also no evidence for a systematic trend in the residual, which would have pointed to the existence of confounding variables. Once again, while our results indicate that the \ac{BH} ringdown signal in prompt-collapse \ac{NS} mergers depends on the binary parameters, we caution the reader that our fit will need to be refined using higher resolution simulation data, before it can be applied in \ac{GW} astronomy.

\newtxt{The fact that $A_{21}/A_{22}$ depends primarily on $q$ and $\tilde\Lambda$ is not unexpected. It is well known that $\tilde\Lambda$ encodes matter effects to leading order in the inspiral of \ac{NS} binaries \cite{Favata:2013rwa}. Moreover, numerical relativity simulations have shown that there are quasi-Universal relations relating $\tilde\Lambda$ to the \ac{GW} amplitude, as well as to the binding energy and angular momentum of the binaries at merger \cite{Read:2013zra, Bernuzzi:2014kca}. These relations were previously known to extend also in the postmerger, if the remnant is a rotating massive \acp{NS} \cite{Bernuzzi:2015rla, Takami:2014tva}. Our work shows that $\tilde\Lambda$ continues to be the most relevant parameter encoding matter effects even after \ac{BH} formation.}

\section{Conclusions}
\label{sec:conclusions}

In this paper, we have studied the \ac{GW} signal produced by the ringdown of the final \ac{BH} formed in prompt-collapse \ac{NS} mergers. To this aim, we analyzed the data of \newtxt{47} numerical relativity simulations and extracted amplitudes and complex frequencies corresponding to the $(\ell=2,m=2)$ and $(\ell=2,m=1)$ modes. While the complex frequencies are expected to depend only on the properties of the remnant \ac{BH}, the amplitudes can encode information of the non-linear dynamics leading to its formation. In particular, we have analyzed the dependency of the amplitudes on the binary mass ratio $q$ and tidal parameter $\tilde\Lambda$.

We find no evidence of a dependency of the amplitude of the $(\ell=2,m=2)$ mode on the binary parameters. However, our data shows that the ratio of the $(\ell=2,m=1)$ and $(\ell=2,m=2)$ mode amplitudes, $A_{21}/A_{22}$, depends on both $q$ and $\tilde\Lambda$. \newtxt{We cannot exclude that other \ac{EOS} parameters other than $\tilde\Lambda$ can influence $A_{21}/A_{22}$, but this appears unlikely given that our dataset includes a large variety of \acp{EOS} and binaries with different \ac{EOS}, but similar $\tilde\Lambda$ and $q$, show similar values of $A_{21}/A_{22}$.} Unfortunately, the estimated finite resolution error on $A_{21}/A_{22}$ is at the level of \newtxt{37\%}, preventing us from fully quantifying this dependency. We nevertheless present a simple phenomenological fit to demonstrate the leading order trends, while cautioning the reader that more accurate numerical relativity data will be needed before such fits can be used for \ac{GW} astronomy applications. \newtxt{Our results show that, while the final \ac{BH} formed in binary \ac{NS} mergers is only characterized by its mass and spin, in accordance with the no-hair theorem \cite{Israel1967-is, Carter1971-we}, the amplitude of the ringdown modes depends on the properties of the progenitor binary. In other words, we can confidently say that \acp{BH} ``know'' about the properties of matter ($\tilde\Lambda$) that formed them.} 

\newtxt{Our results show that $\tilde\Lambda$ can be determined independently from either the inspiral, or the ringdown signals, for sufficiently loud events. The requirement of consistency between the two estimates of $\tilde\Lambda$ could be used to perform null tests of general-relativity in the presence of matter. Such test would not require beyond-general-relativity models of the ringdown, but it would require more accurate numerical relativity data. Assuming Gaussian noise, the ratio $A_{21}/A_{22}$ could be determined with a relative precision of ${\sim}1/\mathrm{SNR}$, where SNR is the signal to noise ratio in the quieter of the two modes. This estimate also assumes that the inclination of the source can be independently measured with sufficient accuracy, which would be the case if an electromagnetic counterpart to the merger can be found. Since the numerical uncertainty is at the level of $37\%$, systematic errors in the estimate of $\tilde\Lambda$ from the ringdown would already dominate over statistical errors for $\mathrm{SNR} \gtrsim 3$.} Performing the required simulations will be the objective of our future work. \newtxt{We note that simulations with grid spacing more than 10 times finer than the ones presented here have already been published \cite{Kiuchi:2017pte}. As such, estimates of the ringdown amplitudes with precision $10{-}100$ times better than the ones reported here are fasible with current simulation technology.}

\section*{Acknowledgements}
DR acknowledges funding from the U.S. Department of Energy, Office of Science, Division of Nuclear Physics under Award Number(s) DE-SC0021177, DE-SC0024388, and from the National Science Foundation under Grants No.~PHY-2011725, PHY-2020275, PHY-2116686, and AST-2108467.
The work of APe is partially funded by the European Union under NextGenerationEU. PRIN 2022 Prot. n. 2022KX2Z3B.

NR simulations were performed on Bridges, Comet, Stampede2 (NSF XSEDE allocation TG-PHY160025), NSF/NCSA Blue Waters (NSF AWD-1811236) supercomputers. Computations for this research were also performed on the Pennsylvania State University’s Institute for Computational and Data Sciences’ Roar supercomputer.
The authors acknowledge PRACE for awarding them access to Joliot-Curie at GENCI@CEA (project: 2019215202, allocation RA5202).
Computations were also performed on the supercomputer SuperMUC-NG at the Leibniz-Rechenzentrum Munich, and on the national HPE Apollo Hawk at the High Performance Computing Center Stuttgart (HLRS). The authors acknowledge the Gauss Centre for Supercomputing e.V. (\url{www.gauss-centre.eu}) for funding this project by providing computing time to the GCS Supercomputer SuperMUC-NG at LRZ (allocation {\tt pn68wi}). The authors acknowledge HLRS for funding this project by providing access to the supercomputer HPE Apollo Hawk under the grant number {\tt INTRHYGUE/44215}.

\section*{References}
\bibliographystyle{iopart-num}
\providecommand{\newblock}{}

\acrodef{AIC}{accretion-induced collapse}
\acrodef{ADM}{Adaptive deconvolution model}
\acrodef{ADM}{Arnowitt-Deser-Misner}
\acrodef{AMR}{adaptive mesh-refinement}
\acrodef{BH}{black hole}
\acrodef{BBH}{binary black-hole}
\acrodef{BHNS}{black-hole neutron-star}
\acrodef{BNS}{binary neutron star}
\acrodef{CCSN}{core-collapse supernova}
\acrodefplural{CCSN}[CCSNe]{core-collapse supernovae}
\acrodef{CMA}{consistent multi-fluid advection}
\acrodef{CFL}{Courant-Friedrichs-Lewy}
\acrodef{DG}{discontinuous Galerkin}
\acrodef{HMNS}{hypermassive neutron star}
\acrodef{EM}{electromagnetic}
\acrodef{ET}{Einstein Telescope}
\acrodef{EOB}{effective-one-body}
\acrodef{EOS}{equation of state}
\acrodef{FF}{fitting factor}
\acrodef{FRB}{fast radio burst}
\acrodef{GR}{general-relativistic}
\acrodef{GRB}{gamma-ray burst}
\acrodef{GRLES}{general-relativistic large-eddy simulation}
\acrodef{GRHD}{general-relativistic hydrodynamics}
\acrodef{GRMHD}{general-relativistic magnetohydrodynamics}
\acrodef{GW}{gravitational wave}
\acrodef{KH}{Kelvin-Helmholtz}
\acrodef{KHI}{Kelvin-Helmholtz instability}
\acrodef{ILES}{implicit large-eddy simulations}
\acrodef{LIA}{linear interaction analysis}
\acrodef{LES}{large-eddy simulation}
\acrodefplural{LES}[LES]{large-eddy simulations}
\acrodef{MHD}{ magnetohydrodynamics}
\acrodef{MRI}{magnetorotational instability}
\acrodef{NR}{numerical relativity}
\acrodef{NS}{neutron star}
\acrodef{QNM}{quasi-normal mode}
\acrodef{PNS}{protoneutron star}
\acrodef{RMNS}{remnant massive neutron star}
\acrodef{SASI}{standing accretion shock instability}
\acrodef{SPH}{smoothed particle hydrodynamics}
\acrodef{SN}{supernova}
\acrodefplural{SN}[SNe]{supernovae}
\acrodef{SNR}{signal-to-noise ratio}
\acrodef{ZAMS}{zero age main sequence}

\end{document}